\documentstyle[twocolumn,aps,epsfig,floats]{revtex}
\begin{document}
\twocolumn[\hsize\textwidth\columnwidth\hsize\csname %
@twocolumnfalse\endcsname
\title{Spin and charge inhomogeneities in  
 high-$T_c$ cuprates: \\
 Evidence from NMR and neutron scattering  experiments
}
\author{Dirk K. Morr $^{1}$, J\"{o}rg Schmalian $^{2}$, 
and David Pines $^{1,3}$}
\address{ $^{1}$ University of Illinois at Urbana-Champaign,  
Loomis Laboratory of Physics, 1110 W. Green St., Urbana, IL 61801\\
$^{2}$ISIS Facility, Rutherford Appleton Laboratory, 
                 Chilton, Didcot,  Oxfordshire, OX11 0QX United Kingdom  \\
$^{3}$ Institute for Complex Adaptive Matter, University of California, and 
LANSCE-Division, Los Alamos National Laboratory, Los Alamos, NM 87545}
\date{\today}
\draft
\maketitle
\begin{abstract}
In this communication we consider the doping dependence of the strong 
antiferromagnetic spin fluctuations in the cuprate superconductors. 
 We investigate the effect of an 
incommensurate magnetic response, as recently observed in inelastic
 neutron scattering  (INS) experiments on 
several 
YBa$_2$Cu$_3$O$_{6+x}$ compounds, on the spin-lattice and spin-echo 
relaxation rates measured in nuclear magnetic resonance (NMR) experiments. We conclude that a consistent
theoretical description of INS and NMR can be reached if one assumes   
spatially 
  inhomogeneous but  locally commensurate   spin correlations and that  NMR and INS experiments can be   described 
within a single theoretical scenario. We discuss a 
simple scenario of spin and charge  inhomogeneities which  includes the main physical 
ingredients 
required for consistency  with experiments.  
\end{abstract}

]

\narrowtext

\section{Introduction}
\label{intro}
Understanding the doping, frequency and temperature dependence of the magnetic
 response in the high-$T_c$ cuprates is one of the most challenging 
problems for the high-$T_c$ community.
 Insight into the nature of the strong antiferromagnetic fluctuations  
very likely holds the key to the unusual normal state properties of the 
cuprates \cite{general,DJS}, 
as seen in a variety of experimental techniques. 
Both nuclear magnetic resonance (NMR) and inelastic neutron 
scattering (INS) experiments are  important tools to probe spin 
excitations in cuprates \cite{NMRreview,Mason1,Bourges1}. 
While INS experiments provide insight into the momentum 
and frequency resolved imaginary part of the spin susceptibility, $\chi({\bf 
q},\omega)$, 
NMR experiments yield information on the momentum averaged 
real and imaginary part of $\chi({\bf q},\omega)$ in the zero frequency limit.
 However, in contrast to NMR experiments, INS measurements rely on the 
existence 
of large single crystals and often
 suffer from a rather  limited  experimental resolution. 
Therefore, these experimental techniques are complementary 
in the information provided on spin excitations in the cuprates. 

One of the central questions in the cuprate superconductors is whether 
INS and NMR experiments can be simultaneously understood 
within a single theoretical scenario.  
Thus far it is not even been clear whether  one can  reach agreement
between INS and NMR data as far as the order of magnitude of the 
spin susceptibility  is concerned. This problem is caused, in part, by the fact
that NMR is sensitive to all magnetic fluctuations, 
regardless of whether they are
related to a pronounced momentum dependence of the spin 
susceptibility.
 In contrast, INS typically probes only those
parts of the susceptibility which are strongly momentum dependent;
the rest are typically attributed to the ``background" of the signal,
as might be caused  by the scattering of neutrons on nuclei and phonons.

The situation has been further complicated by recent  INS experiments on  
YBa$_2$Cu$_3$O$_{6+x}$ which observe  a crossover from a substantial 
incommensuration in the magnetic response at low frequencies, to a commensurate 
structure in $\chi$ at higher frequencies \cite{Tra92,Dai97,Dai98}. The 
question 
thus arises: does this incommensurate order 
originate from spatially inhomogeneous correlations 
between the doped holes \cite{FN1}, which would preserve a locally commensurate 
magnetic response, or does it reflect  homogeneous  incommensuration,
 most likely  due to Fermi surface effects. 

In this communication we argue that since NMR experiments probe the local 
spin environment around a nucleus, they are able to distinguish between a 
locally commensurate and incommensurate magnetic response. 
Earlier theoretical studies of NMR experiments on 
YBa$_2$Cu$_3$O$_{6+x}$ and YBa$_2$Cu$_4$O$_8$ used a phenomenological form of 
$\chi$ \cite{MMP90}:
\begin{equation}
\chi({\bf q},\omega) = { \alpha \xi^2 \over 1+\xi^2 ( {\bf q}-{\bf Q})^2 
-i \omega/\omega_{sf} }  \, , 
\label{chi}
\end{equation}
where $\xi$ is the magnetic correlation length in units of the lattice 
constant, 
$a_0$, $\omega_{\rm sf}$ an energy
scale characterizing the diffusive spin excitations,  $\alpha$ an overall 
temperature independent constant, and ${\bf Q}$  is the position of the
peak in momentum space which was assumed to be commensurate, i.e. ${\bf 
Q}=(\pi,\pi)$.
Using this expression for $\chi({\bf q},\omega)$ a rather detailed quantitative
understanding of various NMR data has been  reached \cite{Zha96,Sok93,BP95}.
In particular, from the analysis of the longitudinal spin lattice relaxation 
rate
of the $^{63}$Cu nuclei, $1/^{63}T_1$, and the spin spin relaxation time, 
$1/T_{\rm 2G}$, 
scaling laws like $\omega_{\rm sf} \propto \xi^{-z}$ with dynamical critical 
exponent, $z$,
 have been  deduced. It was found that $z\approx 1$ between a lower crossover 
temperature, $T_*$, 
and a higher one, $T_{cr}$, whereas $z\approx 2$ above $T_{cr}$. In the 
temperature range
 where $z=1$ scaling applies, it follows that $T_1T/T_{2G}$ is independent of
 temperature; a conjecture which was experimentally verified  
by Curro {\it et al.} for YBa$_2$Cu$_4$O$_8$ \cite{Cur97}.
Below the pseudogap temperature, $T_*$, $\omega_{\rm sf}$ and 
$\xi$ decouple and a quasiparticle and spin pseudogap  emerges. 
Moreover, it was recently shown that the temperature dependence of 
 the incommensurate peak width,  as determined in  
 INS experiments on La$_{2-x}$Sr$_x$CuO$_4$ \cite{AMH97},  
 showed $z=1$ scaling  over a wide range of temperatures, in agreement with the 
NMR findings for YBa$_2$Cu$_4$O$_8$.
Because of its appearance in both NMR and INS results, we will therefore assume  in
the following analysis  that $z=1$ is the
proper scaling behavior for magnetically underdoped cuprates between $T_*$ and 
$T_{\rm cr}$.

One of our central results is that the attempt to understand the NMR data for 
YBa$_2$Cu$_4$O$_8$ with homogeneous incommensuration is inconsistent with  $z=1$ 
scaling. This implies that the local magnetic response, which is probed in NMR 
experiments, is commensurate. Should INS experiments show that this compound 
exhibits a (globally) incommensurate magnetic response similar to 
that seen in 
La$_{2-x}$Sr$_x$CuO$_4$, this  would strongly support a dynamic charge and spin inhomogeneity (stripe) origin of 
the incommensuration. In Sec.~V we discuss a {\it spin and charge inhomogeneity} 
scenario which reconciles the global incommensuration with the local 
commensuration in $\chi$.

In a second important result we carry out a quantitative comparison of the strength of the 
antiferromagnetic spin fluctuations, as measured by NMR and INS experiments.
We find that agreement between the results obtained from these quite different experimental techniques, which explore not only a different frequency range, but different wavevector regimes, can be obtained within a factor of 2. Given the large uncertainties in 
determining the absolute value of $\chi''$, we believe that this result 
demonstrates that a consistent description of INS and NMR data can be achieved 
within the framework of Eq.(1).

Although the form of $\chi$ in  Eq.(\ref{chi})  was originally invented  to 
understand the spin response  at  very low frequencies
and above the superconducting transition temperature \cite{MMP90}, it has proved 
interesting to investigate to what extent one can understand INS data at higher 
frequencies within the same framework.  
The results by Aeppli {\em et al.}  support the picture of  a unique  
incommensurate spin 
response   from  zero energy  up to  $15\, {\rm meV}$, the highest  energy used 
in Ref.\cite{AMH97}.
Within the error bars of the experiment, only for momentum values away from the 
peak maximum do systematic deviations from Eq.(\ref{chi}) occur.
Such deviations indicate the presence of lattice corrections to  the continuum 
limit
used  in Eq.(\ref{chi}). Lattice corrections are expected to be extremely
 important for  the local, momentum averaged, susceptibility:
\begin{equation}
\chi_{loc}''(\omega)= {1 \over 4 \pi^2} \int_{BZ} d^2 {\bf q}  \, \chi''({\bf 
q},\omega) \,  ,
\label{chiloc}
\end{equation}
since the phase space of momenta away from the peak maxima is considerable in 
two dimensions. For higher frequencies spin excitations away from the 
antiferromagnetic peak and energetically large compared to $\omega_{\rm sf}$ 
come into play and we therefore expect a poor description of 
$\chi''_{loc}(\omega)$ in terms of Eq.(\ref{chi}). This conclusion is 
independent of whether the peak position commensurate or incommensurate 
peaks. 

Deviations from the universal continuum limit of $\chi$ are expected to be also 
of 
  relevance for the relaxation rates measured in NMR 
experiments since these are weighted momentum averages of 
$\chi({\bf q},\omega)$.
Thus, it is worthwhile to study
whether one can find  indications for lattice corrections from an 
analysis of NMR and INS data.
Such corrections to $\chi$ are also of relevance
for our understanding of the lifetime of so-called {\em cold} quasiparticles if 
one assumes that the lifetime of these quasiparticles is also dominated by 
scattering off spin fluctuations \cite{HR,SP96,IM98,Sto_pc}.

Our paper is organized as follows. 
In Sec.~\ref{theory} we give a brief overview of the 
theoretical framework in which we analyze the INS and NMR data. 
In Sec.~\ref{1248} we analyze NMR data on YBa$_2$Cu$_4$O$_8$ for both 
commensurate and incommensurate magnetic response and discuss the 
role of lattice corrections. In Secs.~\ref{1237} and \ref{1236} we analyze NMR 
data 
on two YBa$_2$Cu$_3$O$_{6+x}$ compounds. In Sec.~\ref{INS} we 
discuss the consistency between NMR and INS data, as well as the role 
of lattice corrections for the local susceptibility. 
In Sec.~\ref{mag_stripes} we propose a {\em spin and charge inhomogeneity} scenario as 
a 
possible way
to reconcile NMR and INS data. Finally, in Sec.~\ref{concl} we draw our 
conclusions.

\section{Theoretical Overview}
\label{theory}

We briefly discuss the theoretical framework 
in which we discuss NMR and INS experiments. 
In order to analyze the low frequency NMR data,  we use the 
dynamical spin susceptibility of Eq.(\ref{chi}), where for the commensurate 
case ${\bf Q}=(\pi,\pi)$. To allow for an incommensurate structure of $\chi$, 
we use Eq.(\ref{chi}) with ${\bf Q}=(1,1 \pm \delta)\pi$ and ${\bf Q}=(1 \pm 
\delta,1)\pi$ and sum  over all four peaks \cite{Zha96}.
The inclusion of the correct form of lattice corrections is rather difficult 
since it requires a microscopic model which is 
 beyond the scope of this paper. 
In general we expect lattice corrections to appear in the form of an
upper momentum cutoff, $\Lambda$.
In the following we choose a soft cutoff procedure for $\delta {\bf q} \equiv 
{\bf q}-{\bf Q}$ in the denominator of Eq.(\ref{chi}):
\begin{equation}
\delta {\bf q}^2 \rightarrow 
\delta {\bf q}^2 \left(1+  { \delta {\bf q}^2 \over \Lambda^2 } \right)    \, .
\label{lattice}
\end{equation}
We also expect a weak momentum dependence of
$\alpha$ and $\omega_{\rm sf}$ once the continuum description
breaks down; however,  to keep the number of tunable parameters  small we 
ignore these effects.

In NMR experiments,  one measures the spin-lattice relaxation rate $1/T_{1x}$, 
with applied magnetic 
field in $x$-direction,
 and the spin-echo  rate $1/ T_{2G}$, which  can be expressed in terms of
the dynamical susceptibility as:
\begin{eqnarray}
{1 \over T_{1x} T} &=& { k_B \over 2 \hbar} (\hbar^2 \gamma_n \gamma_e)^2 
{ 1 \over N} \sum_q F_x(q) \lim_{\omega \rightarrow 0} { \chi''(q, \omega) 
\over 
\omega} \ , \label{T1T} \\
\Big({1 \over T_{2G} }\Big)^2 &=& { 0.69 \over 128 \hbar^2} (\hbar^2 \gamma_n 
\gamma_e)^4 \Bigg\{ { 1 \over N} \sum_q \Big[F_{ab}^{eff}(q) \chi'(q, 
\omega)\Big]^2 \nonumber \\
& & -  \Big[\sum_q F_{ab}^{eff}(q) \chi'(q, \omega)\Big]^2 \Bigg\}
\label{T2G} \ ,
\end{eqnarray}
where $x=ab,c$ describes the direction of the external magnetic field.  
The $^{63}Cu$ form factors are given by 
\begin{eqnarray}
^{63}F_c(q) &=&\Bigg[A_{ab}+2B \Big( cos(q_x)+cos(q_y) \Big) \Bigg]^2 \ , \\
^{63}F_{ab}^{eff}(q)&=&\Bigg[A_c+2B \Big( cos(q_x)+cos(q_y) \Big) \Bigg]^2 \ , 
\\
^{63}F_{ab}(q)&=&{1 \over 2} \Big[ ^{63}F_{ab}^{eff}(q) + ^{63}F_c(q) \Big] \ .
\end{eqnarray}
where $A_{ab}, A_c$ and $B$ are the on-site and transferred hyperfine coupling 
constants, respectively \cite{Zha96}.

It follows from Eq.(\ref{chi}) that the spin excitations in the normal state 
are 
completely described by three  parameters, $\alpha, \xi$, and $\omega_{sf}$. In 
order to extract these parameters from the experimental NMR data, we also need 
to obtain the three hyperfine coupling constants, 
$A_{ab}, A_c$, and $B$; hence we have six unknown parameters in the above 
equations, and 
 require six equations to determine them. So far we have two, Eqs.(\ref{T1T}) 
and (\ref{T2G}). 
An additional constraint arises from the temperature independence of the 
$^{63}$Cu Knight shift
in a magnetic field parallel to the $c$ axis in YBa$_2$Cu$_3$O$_7$ 
\cite{Zha96} which yields
\begin{equation}
A_c+4B \approx 0 \ .
\label{constr1}
\end{equation}
A fourth constraint comes from the anisotropy of the $^{63}$Cu spin-lattice 
relaxation 
rates \cite{Barrett}, which
for YBa$_2$Cu$_3$O$_{7}$ was measured to be 
\begin{equation}
^{63}R ={T_{1c} \over T_{1ab} } \approx 3.7 \pm 0.1 \ .
\label{constr2}
\end{equation}
A fifth constraint involving the hyperfine coupling constants can be obtained 
by 
 plotting the Knight shift  $^{63}K_{ab}$ versus $\chi_0(T)$ \cite{Zha96}, 
which 
yields
\begin{equation}
4B+A_{ab} \approx 200 { kOe \over \mu_B} \ .
\label{constr3}
\end{equation}
A final constraint is obtained from the earlier conjecture that the
antiferromagnetic correlation length at the crossover temperature 
$T_{\rm cr}$ is approximately two lattice constants \cite{BP95}. 
Recent microscopic calculations have confirmed this conjecture, 
showing that indeed $\xi(T_{cr}) $ is of the order of a few lattice 
constants\cite{SPS97}.

To the extent that  experimental data for $T_1$ and $T_{2G}$ are available 
up to $T_{cr}$, one can fit the above set of six equations 
self-consistently to the data, and thus obtain not only the temperature 
independent 
parameters $\alpha, A_{ab}, A_c$, and $B$, 
but also the temperature dependence of $\xi$ and $\omega_{sf}$. It turns out, 
however, that this analysis is only possible for  
YBa$_2$Cu$_4$O$_8$, since this is the single compound for which data up to 
sufficiently high temperatures have been obtained \cite{Cur97}. 
Our corresponding results will be presented in section~\ref{1248}.
In order to extract the relevant parameters from NMR data on the related,  
widely 
studied, YBa$_2$Cu$_3$O$_{6+x}$ compounds, and thus to make a comparison 
between 
NMR and 
INS experiments possible,  we need to make two assumptions, which, at least 
partly, will be supported by the experimental data for  YBa$_2$Cu$_4$O$_8$. 
 First, we assume a  relation between $\omega_{sf}$ and $\xi$,   given by 
\cite{Sok93,BP95}
\begin{equation}
\omega_{sf}  = { \hat{c} \, \xi^{-z} } \ ,
\label{scale1}
\end{equation}
where $\hat{c}$ is a temperature independent constant.  
For YBa$_2$Cu$_4$O$_8$, where we can independently extract $\omega_{sf}$ and 
$\xi$, we will show that there is indeed a crossover from 
$z=1$ behavior at low temperature to $z=2$ behavior at higher temperatures.
The second assumption concerns the   temperature dependence of the magnetic 
correlation length, $\xi(T)$, which we take to be that obtained by one of us 
using a renormalization group (RG) approach \cite{Sch98}: 
for $z=1$  the temperature dependence of the magnetic correlation length 
is given by
\begin{equation}
\xi^{-2} = \xi_0^{-2} + b T^2 \ ,
\label{xi}
\end{equation}
a result which is in good agreement with INS experiments on 
La$_{2-x}$Sr$_x$CuO$_4$ \cite{AMH97}. 
Having all the necessary theoretical tools in place, we now address 
the questions raised in the introduction.

\section{Analysis of NMR experiments}

\subsection{YBa$_2$Cu$_4$O$_8$}
\label{1248}

YBa$_2$Cu$_4$O$_8$ is a benchmark system for our analysis, 
since Curro {\it et al.} \cite{Cur97} have measured the 
relaxation times $T_1$ and $T_{2G}$ over a wide temperature range from 80K to 
750K, and in particular between $T_*approx 200$ K and $T_{cr}approx 500$ K.
We are therefore able to extract the relevant parameters which we discussed 
above from a self-consistent fit of Eqs.(\ref{T1T})- (\ref{constr3}) to the 
$T_1$ and $T_{2G}$ data. In doing so, we assume that the constraints given by 
Eqs.(\ref{constr1})-(\ref{constr3}) for YBa$_2$Cu$_3$O$_7$ are also valid for 
YBa$_2$Cu$_4$O$_8$. 
Since we can determine $\xi$ and $\omega_{sf}$ independently, we are able to 
study the effect of both an incommensurate magnetic response, and of 
non-universal 
lattice corrections, on the scaling law $\omega_{sf}  = \hat{c}/\xi^z $.

\begin{figure} [t]
\begin{center}
\leavevmode
\epsfxsize=7.5cm
\epsffile{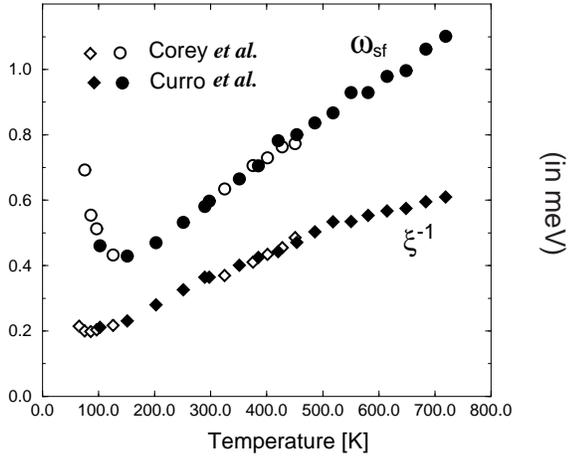}
\end{center}
\caption{ The temperature dependence of  $\omega_{sf}$ and 
 $\xi^{-1}$ in YBa$_2$Cu$_4$O$_8$ for the commensurate case with $\Lambda=2 
\pi^{1/2}$.}
\label{temp}
\end{figure}

\begin{figure} [t]
\begin{center}
\leavevmode
\epsfxsize=7.5cm
\epsffile{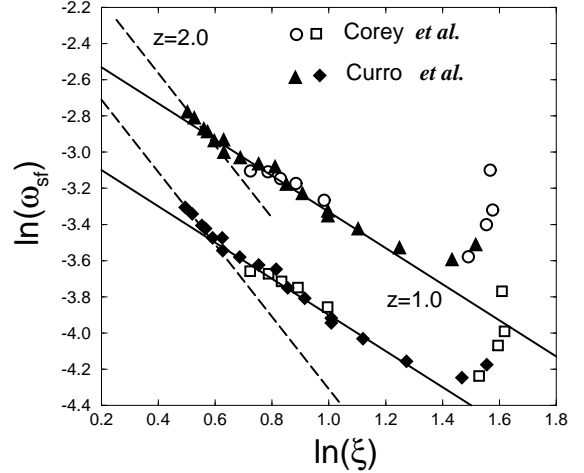}
\end{center}
\caption{$ln(\omega_{sf})$ as a function of $ln(\xi)$ for the commensurate 
case. 
Upper 
data set: $\Lambda=2 \pi^{1/2}/4$; 
Lower data set: $\Lambda=2\pi^{1/2}$. Solid line: z=1; dashed line: z=2.}
\label{omsf_xi}
\end{figure}

We first reconsider the case of a commensurate structure of $\chi$, 
and set $\Lambda=2\sqrt{\pi}$, i.e. we choose a   momentum cut-off which 
corresponds to the linear size  of the BZ.
 We present the temperature dependence of both $\omega_{sf}$ and $\xi^{-1}$, 
 which results from a self-consistent fit to the experimental data, in 
Fig.~\ref{temp}.
Between $T_*=200$ K and $T_{cr}=500$ K, $\omega_{sf}$ and $\xi^{-1}$ clearly 
scale linearly with  temperature. In order to study the extent to which 
$\omega_{sf}$ and $\xi$ obey the above scaling law, we plot in 
Fig.~\ref{omsf_xi}, $\ln(\omega_{sf})$ as a function of $\ln(\xi)$ (lower 
curve). 
 The dynamical scaling range is of course too limited to
  prove  the existence of a scaling relation. However, assuming $\omega_{\rm 
sf} 
\propto \xi^{-z}$,
we  find  $z\approx 1$ below $500\, {\rm K}$, which we identify with $T_{\rm 
cr}$
and $z\approx 2$ above $T_{\rm cr}$. 
These results are consistent with an earlier analysis of the scaling 
behavior~\cite{Sok93,BP95} and the microscopic 
scenario of Ref.~\cite{SPS97}.
To study the effect of stronger lattice corrections, 
we decreased $\Lambda$ to  $\Lambda=2\sqrt{\pi}/4$ (upper curve in 
Fig.~\ref{omsf_xi} which for clarity is offset).  Our conclusions concerning 
the scaling 
relations remain unchanged.
We can thus conclude that  the dynamical scaling behavior of   spin excitations 
is robust against even sizeable lattice corrections, as long as the structure 
of 
$\chi''$ is commensurate.
\begin{figure} [t]
\begin{center}
\leavevmode
\epsfxsize=7.5cm
\epsffile{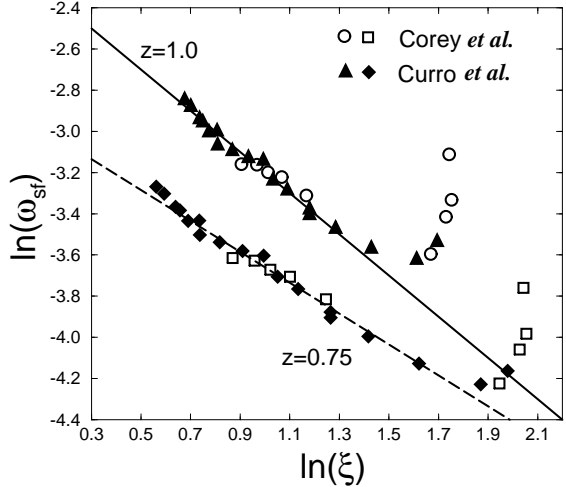}
\end{center}
\caption{$ln(\omega_{sf})$ as a function of $ln(\xi)$ for the incommensurate 
case.
Lower data set: $\Lambda=2 \pi^{1/2}$; 
Upper data set: $\Lambda=2\pi^{1/2}/15$. Solid line: z=1; dashed line: z=0.75.}
\label{omsf_xi_inc}
\end{figure}

We next examine the effect of incommensuration on the dynamical scaling. It was 
earlier argued that
the magnetic response in  YBa$_2$Cu$_4$O$_8$ should be very similar
 to YBa$_2$Cu$_3$O$_{6.8}$~\cite{BP95}. Using the observed doping dependence 
of the incommensuration in YBa$_2$Cu$_3$O$_{6+x}$, we estimate the 
incommensuration in YBa$_2$Cu$_4$O$_8$
to be $\delta=0.23$. Setting $\Lambda=2\sqrt{\pi}$ we plot in 
Fig.~\ref{omsf_xi_inc}  $\ln(\omega_{sf})$ as a function of $\ln(\xi)$ (lower 
curve) and find, following the same argumentation as above,
 that the dynamical scaling exponent below $T_{\rm cr} \approx 500$ K is now 
$z\approx 0.75$. 
 Increasing the strength of the lattice corrections by decreasing $\Lambda$ we  
find  that $z$   
is increased. However,  in order to obtain again $z \approx 1$  
(the upper curve in Fig.~\ref{omsf_xi_inc}), which one would expect from the  
INS data on La$_{2-x}$Sr$_x$CuO$_4$, we need a very small
 momentum cut-off, $\Lambda \approx 2 \sqrt{\pi}/ 15$. Such a small cut-off  
of order $O(1/\xi)$ is inconsistent with the continuum theory which
is the basis for a dynamical scaling  approach.
We therefore conclude that a spatially  homogeneous  incommensurate 
magnetic response is in
 contradiction with  the available NMR data.
The physical origin of this sensitivity of the magnetic response of a 
homogeneously incommensurate
system, compared to a locally commensurate one, results from the fact that the 
former effectively
decreases the spatial extent of the spin-spin
correlations, thus increasing the role of large values of $\delta{\bf q} \sim 
\Lambda$.
It should also be noted that our arguments using lattice 
corrections to discriminate between these two scenarios works only for 
intermediate values of the correlation length.
 For very large $\xi$ the system should be insensitive to the cut-off procedure 
regardless of
whether it is commensurate or incommensurate.

 In what follows we assume that the low frequency magnetic response measured in 
an NMR experiment is indeed locally commensurate which implies that $z=1$ 
scaling prevails between $T_*$ and $T_{cr}$. In Sec.~\ref{mag_stripes}  we 
present a possible theoretical scenario to resolve this apparent contradiction 
between  
the incommensuration seen in INS experiments and our conclusions
based on the available NMR data.
Finally, using the commensurate form of Eq.(\ref{chi}), 
and a momentum cut-off $\Lambda= 2 \sqrt{\pi}$ we find the following 
parameters from the solution of the above self-consistent equations:
$\alpha= 10 eV^{-1}$, $A_{ab}=20.2\, kOe/\mu_B$, $A_{c}=-182.8\, 
kOe/\mu_B$, and  $B=45.7 \, kOe/\mu_B$ for YBa$_2$Cu$_4$O$_8$. 

\subsection{YBa$_2$Cu$_3$O$_7$}
\label{1237}

Since it was earlier argued that for the optimally doped YBa$_2$Cu$_3$O$_7$, 
$T_{cr} 
\approx 125$ K, only slightly above $T_c$,  there is no  significant 
temperature 
dependence of the relaxation rates between  $T_c$ and $T_{cr}$. We will 
therefore perform our
 analysis of the NMR data only at $T_{cr}$.  Assuming that $\xi(T_{cr} ) = 2$, 
and setting 
$\Lambda=2\sqrt{\pi}$, 
we utilize $T_{2G}$ data by Itoh {\it et al.}~\cite{Itoh94} and Stern {\it et 
al.}~\cite{Stern95}, 
as well as $T_{1c}$ data by Hammel {\it et al.}~\cite{Ham89} to find  the 
following hyperfine coupling constants $A_{ab}=27.8 \, kOe/\mu_B$, 
$A_{c}=-175.2\, kOe/\mu_B$, and  $B=43.9 \, kOe/\mu_B$. Note that these values 
agree well with the hyperfine coupling constants we extracted for 
YBa$_2$Cu$_4$O$_8$.
With $1/T_{2G} = 10^{4}$ s$^{-1}$ and $T_{1c}T \approx 0.15$ Ks at $T_{cr}$,  
we 
obtain $\alpha=18.5$ eV$^{-1}$, 
$\omega_{sf} \approx 19.8 $ meV  and  $\hat{c} \approx 39.6$ meV. 
It turns out that the above parameter set is  rather robust against changes in 
$\Lambda$.  
When $\Lambda$ is decreased from  $\Lambda=2\sqrt{\pi}$ to 
$\Lambda=2\sqrt{\pi}/4$, $A_c$ and $B$ 
decrease by about 1.5\%, $A_{ab}$ increases by about 8.5\%, $\alpha$ increases 
by 11\%, 
and $\omega_{sf}$ decreases by about 15\%. A change of the momentum cut-off by 
a 
factor of 4, 
which   decreases the area of integration in the BZ by a factor of 16, thus 
leads 
only to moderate changes in the parameter set. However, as we show in 
Sec.~\ref{INS}, the corresponding
 changes in the  local susceptibility at high frequencies are much more 
dramatic.

\subsection{YBa$_2$Cu$_3$O$_{6.63}$}
\label{1236}
\begin{figure} [t]
\begin{center}
\leavevmode
\epsfxsize=7.5cm
\epsffile{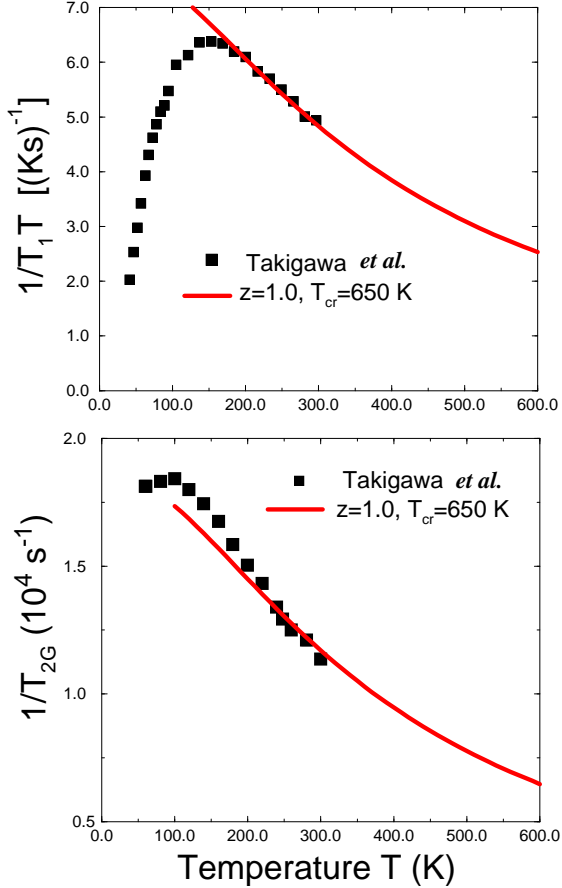}
\end{center}
\caption{Comparison of our theoretical results for the temperature dependence 
of 
$1/T_1T$ (upper figure), and $1/T_{2G}$ (lower figure) in 
YBa$_2$Cu$_3$O$_{6.63}$ 
with the 
experimental data by Takigawa {\it et al.}~\protect\cite{Tak91}.}
\label{NMR_O663}
\end{figure}

In what follows we show that, though only a limited set of NMR data on 
YBa$_2$Cu$_3$O$_{6.63}$
is available, we nevertheless reach the same conclusions regarding the scaling 
behavior  as in 
YBa$_2$Cu$_4$O$_8$.
In particular, since no NMR data on YBa$_2$Cu$_3$O$_{6.63}$ are available above 
$T=300$ K, and since we do not know the exact value of $T_{cr}$, we cannot 
extract the  parameter set  from a self-consistent fit to the NMR data as
 we did in the case of YBa$_2$Cu$_4$O$_8$. 
However, it is in this doping range that INS 
data are available and therefore a consistency check between the 
parameter sets extracted from NMR and INS measurements might be the most  
promising.
It turns out that though we are not able to perform a fully self-consistent 
fit, 
we can still extract the parameter set if we make several additional 
assumptions 
which
 are supported by our results on  YBa$_2$Cu$_4$O$_8$.
First, we assume that $z=1$ scaling is present between $T_*$ and $T_{cr}$ and 
that we can describe the temperature dependence of $\xi$ and the relation 
between $\omega_{sf}$ and $\xi$ by Eqs.(\ref{scale1}) and  (\ref{xi}),   
respectively. 
Second, we assume that the hyperfine coupling constants are only weakly doping 
dependent, 
and that to good approximation we can use the same constants we extracted from 
the analysis of 
YBa$_2$Cu$_3$O$_7$ for  YBa$_2$Cu$_3$O$_{6.63}$. 
Third, we need an estimate for $T_{cr}$, which is unknown for 
YBa$_2$Cu$_3$O$_{6.63}$. 
However, since $T_{cr} \approx 500$ K for 
YBa$_2$Cu$_4$O$_8$, and since $T_{cr}$   increases with decreasing doping,  
we assume $T_{cr} \approx 550-650$ K for YBa$_2$Cu$_3$O$_{6.63}$.  
To demonstrate the effect of the uncertainty in the latter assumption, we 
calculate 
he parameters $\alpha, \omega_{sf}(T)$ and $\xi(T)$ for $T_{cr}=650$ K  and 
$T_{cr}=550$ K as well as for two different values of the momentum cut-off 
$\Lambda$.
 The resulting parameter sets based on experimental data by Takigawa {\it et 
al.} \cite{Tak91} 
are shown in Table~\ref{O663}.  
\begin{table} [t]
\begin{tabular}{ccccc}
  &  $\alpha$ (eV$^{-1}$) & $\hat{c}$ (meV) & $b$ ($10^{-7}$ K$^2$) &  
$\xi_0^{-2}$ ($10^{-2}$) \\[0.2cm] \hline \\
$T_{cr}=650$ K  & 11.0  &  68.1  & 5.1 &  3.5 \\
$\Lambda=2\sqrt{\pi} $ &   &    &  &   \\ \hline \\
$T_{cr}=650$ K  & 12.2  &  61.2  & 5.0 &  3.9 \\
$\Lambda=2\sqrt{\pi}/4$ &   &    &  &   \\ \hline \\
$T_{cr}=550$ K & 13.0  &  67.3  & 6.7 &  4.7 \\
$\Lambda=2\sqrt{\pi}$ &   &    &  &   \\ \hline \\
$T_{cr}=550$ K & 14.5  &  59.9  & 6.5 &  5.3 \\
$\Lambda=2\sqrt{\pi}/4$ &   &    &  &   \\
\hline \\
\end{tabular}
\caption{Parameter sets for two different values of $T_{cr}$ and $\Lambda$.}
\label{O663}
\end{table}
 Note that the parameters in Table~\ref{O663} are rather robust against 
large variations in $T_{cr}$ and/or $\Lambda$ and only differ at the most by 
about 30\%.  
In Fig.~\ref{NMR_O663} we compare our theoretical results for 
 the relaxation rates, $1/T_1$ and $1/T_{2G}$, 
 with the experimental data by Takigawa {\it et al.}~\cite{Tak91}.
We find that the temperature dependence of the relaxation rates calculated for 
each of the four different parameter sets in Table~\ref{O663}, is practically 
indistinguishable; 
we therefore only plot the results for the parameter set with $T_{cr}=650$ K  
and $\Lambda=2\sqrt{\pi}$. 
This result is consistent with the robustness of the scaling behavior in the
 commensurate  case against non-universal lattice corrections as found for 
YBa$_2$Cu$_4$O$_8$.
We find a consistent description 
of the experimental data for  YBa$_2$Cu$_3$O$_{6.63}$  using the assumption of 
a 
$z=1$ scaling
 relationship and local
commensuration in the temperature regime $T_*\approx 230\, {\rm K} < T < 300\, 
{\rm K} $, and, as was the case for YBa$_2$Cu$_4$O$_8$, assuming commensurate 
behavior leads to results which are practically independent of non-universal 
lattice corrections.

\section{Inelastic Neutron Scattering}
\label{INS}

In this section we address the following  two   questions:(i) What is the 
effect 
of lattice corrections on the local susceptibility, and (ii) can one describe 
the available INS data with the parameter set extracted in the previous section 
? 

As we discussed in the introduction we expect lattice corrections  to lead to 
substantial corrections of  $\chi''_{loc}$ at frequencies larger than 
$\omega_{sf}$. Above this energy scale we also expect that the typical energy, 
$\Delta_{\rm sw}$, of  propagating  spin waves,
which, at low frequencies are completely overdamped by particle hole 
excitations, 
comes into play, leading to a modified form of the spin susceptibility 
\begin{equation}
\chi({\bf q},\omega)= { \alpha \xi^2 \over 1 + \xi^2({\bf q}-{\bf Q})^2
- i \omega/\omega_{sf} - (\omega/\Delta_{\rm sw})^2 } 
\label{chiQ} \ .
\end{equation}
For $\omega < \Delta_{\rm sw}^2/\omega_{\rm sf}$ no sign of a propagating peak 
exists due to the diffusive character of the spin excitations.
For $\omega > \Delta_{\rm sw}^2/\omega_{\rm sf}$ and ${\bf q}={\bf Q}$ the 
consequence of
a propagating mode is a  pronounced pole in the excitation spectrum if 
$\Delta_{\rm sw} < \omega_{\rm sf}$ and a   soft   upper cut off in energy
if $\Delta_{\rm sw} > \omega_{\rm sf}$. The form of $\chi({\bf q},\omega)$ in 
Eq.(\ref{chiQ})  
can be shown  to describe the propagating spin mode in YBa$_2$Cu$_3$O$_{6.5}$ 
above $T_c$ \cite{Morr98}.

\begin{figure} [t]
\begin{center}
\leavevmode
\epsfxsize=7.5cm
\epsffile{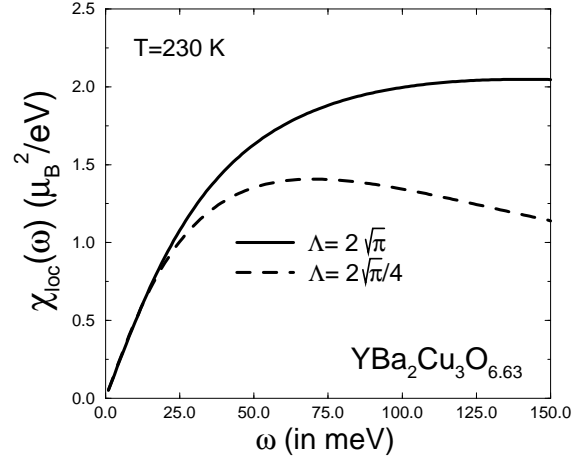}
\end{center}
\caption{The local susceptibility $\chi_{loc}''(\omega)$ as a function 
of frequency for two different values of $\Lambda$.}
\label{chi_loc}
\end{figure}

Using the parameter sets (1) and (2) in Table~\ref{O663}, extracted in the 
previous section from NMR experiments,
we plot in  Fig.~\ref{chi_loc}  the local susceptibility of 
YBa$_2$Cu$_3$O$_{6.63}$ for two different values of $\Lambda$.
\begin{figure} [t]
\begin{center}
\leavevmode
\epsfxsize=7.5cm
\epsffile{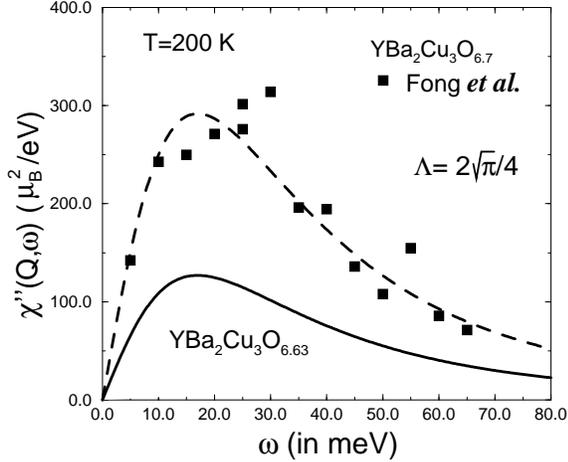}
\end{center}
\caption{$\chi''({\bf Q},\omega)$ for YBa$_2$Cu$_3$O$_{6.63}$. {\it (a)} Solid 
line:  $\chi''({\bf Q},\omega)$ calculated with parameter set (2) in 
Table~\protect\ref{O663} from NMR experiments. {\it (b)} Dashed line: 
$\chi''({\bf Q},\omega)$ from {\it (a)} multiplied by an overall factor of 2.3. 
The experimental data are from Fong {\it et al.}~\protect\cite{Fong96}.}
\label{INS_NMR_com}
\end{figure}
We clearly see that upon decreasing the momentum cut-off, the maximum in 
$\chi''_{loc}$ 
moves towards lower frequencies. Experimentally, however, the intensity at 
higher frequencies 
drops much faster than in Fig.~\ref{chi_loc}, even for 
$\Lambda=2\sqrt{\pi}/4$ \cite{Dai99}. 
In order to explain this discrepancy, we note that at higher frequencies, the 
dominant  contribution to $\chi''_{loc}$ comes from regions in momentum space 
which 
are far away from the peak position at $(\pi,\pi)$. In these regions, 
$\chi''({\bf q}, \omega)$ is only weakly momentum dependent in which case its 
contribution might be easily  attributed
 to the experimental background. In other words, we believe that the origin of 
the 
discrepancy lies in an underestimate of the experimental intensity at
 higher frequencies due to the problems one has in resolving it from the 
background \cite{Dai_pc}.

In Fig.~\ref{INS_NMR_com} we plot the calculated INS intensity, i.e., 
$\chi''({\bf Q}, \omega)$ as a function of  frequency at ${\bf Q}=(\pi,\pi)$. 
Since the calculated intensity is practically the same for all four parameter 
sets in  Table~\ref{O663}, we only present $\chi''({\bf Q}, \omega)$ calculated 
with the second parameter set in Table~\ref{O663} (solid line).
In the same figure, we also include the experimental data by Fong {\it et al.} 
for YBa$_2$Cu$_3$O$_{6.7}$ \cite{Fong96}.
For the temperature range of interest, this material is the closest match for 
YBa$_2$Cu$_3$O$_{6.63}$.
 Calculating the overall INS intensity with the parameters extracted from NMR 
experiments in the previous section, 
we find that we underestimate the experimental INS intensity by about 56 \%, or 
a factor of 2.3 (the dashed line shows the calculated INS intensity, multiplied 
by an overall factor of 2.3). 
Given the uncertainties in the experimental determination of the 
absolute scale of $\chi''$, we believe that the above result represents 
reasonable agreement between the INS and NMR data and thus demonstrates 
consistency in the description of spin excitations based on these two 
experimental techniques.

\section{Spin and Charge Inhomogeneities}
\label{mag_stripes}

We saw in Sec.~\ref{1248} that under the assumption of $z=1$ scaling, supported 
by 
recent INS experiments \cite{AMH97}, a locally incommensurate magnetic response 
is 
inconsistent with the available experimental 
NMR data. We therefore concluded that NMR data 
support a locally commensurate magnetic structure. 
The question thus arises of how one can understand 
a locally commensurate, but globally incommensurate
magnetic response as seen by INS?  It has been suggested 
that charge stripes separated by an average 
distance $l_0=2\pi/\delta$
are the origin of the incommensurate magnetic 
response seen in INS experiments~\cite{stripes}. For a large part
of the sample this would leave 
the locally commensurate structure intact. 
Except for the Nd-doped 214 compounds \cite{tran} and probably for the
particular doping value $x=\frac{1}{8}$ in La$_{2-x}$Sr$_x$CuO$_2$, 
these stripes are believed to be dynamic rather than static in nature. 
 
Considering a spatial charge variation:
\begin{equation}
\rho(r)=\rho_0+\delta\rho(r) \ ,
\end{equation}
where $\rho_0$ is independent of $r$.
The question arises whether the system is in a regime with  $\rho_0 \ll 
\delta\rho(r) $,
as suggested in Ref.~\cite{stripes}, or rather characterized by rather small 
charge
variations, i.e. $\delta\rho(r)< \rho_0 $.
The particular appeal of the former assumption of strong charge modulations is 
an immediate explanation of the doping dependence of the position of the 
incommensurate
peak, $\delta=2x$. Stripes with one hole every second lattice site  in a hole 
free
antiferromagnetic environment move closer together upon doping, leading to the 
above 
$x$-dependence of $\delta$. Also, the  exceptional  behavior of various 
magnetic 
and
transport properties for the doping value $x=\frac{1}{8}$ can be easily 
understood in terms 
of a stable commensurate arrangement of such stripes and the underlying 
lattice.
Another argument in favor of this scenario are the recent results by Vojta and 
Sachdev \cite{Voj99}
who investigated the mean field behavior of a system with competing magnetic 
and long range Coulomb interactions and found areas with a strong charge 
modulation 
separated by regions with barely disturbed antiferromagnetic correlations.

Nevertheless, there are several conceptional problems 
with such a 
strong charge modulation, which lead us to present some arguments favoring a 
more 
moderate
charge variation, $\delta\rho(r)< \rho_0 $,
under circumstances that the transverse  mobility of the charge carriers with 
respect to the averaged
position  of the stripes is large.
 First,  uncorrelated spatial stripes of width $r_0$, which fluctuate  
transversely
over a distance  $d$, will give rise to a 
broadening 
$ d/l_0^2$ of the magnetic peaks observed by INS.  
The experimental INS results, however, provide strong support  for a
scenario in which the width of the magnetic peaks is determined 
by $\xi^{-1}$ \cite{AMH97}.
 Thus,  in order to observe 
separated incommensurate peaks, we need  at least $\xi \gg d$ to establish
a well defined antiphase domain wall,   implying  $d \ll l_0$. 
 In other words, assuming uncorrelated stripes, the width $d$ over which 
stripes 
  fluctuate
 must be very  small    to account for the existing INS data. 
Such "stiff" but uncorrelated stripes seem to be in contradiction to the notion 
of
stripes as a dynamical entity.
Therefore, we arrive at the conclusion that  spatial stripe 
fluctuations, if they exist, must be strongly correlated.

Second, at the characteristic temperatures where stripes with $\rho_0 \ll 
\delta\rho(r) $   
appear, strong modifications   of the resistivity and other  transport  
properties
are expected to occur. None or only   
  moderate changes  of this kind have been observed and 
no indication for a dimensional crossover from quasi one-dimensional
to quasi two-dimensional dynamics seems to be present.

Third, for low frequencies,
the spin excitations  in doped cuprates are overdamped rather than propagating
spin waves. The suppression of the spin damping upon opening
of a quasiparticle gap in the superconducting as well as in the
pseudogap state implies that the dominant source of the spin damping, $\gamma$,
are particle-hole excitations. Assuming weak charge modulations, we then find 
${\rm Im} \chi^{-1}(\omega)  \propto \gamma \omega$, in agreement with the 
results
of INS experiments \cite{AMH97}. 

On the other hand, rigid stripes separated by one dimensional charge carriers, 
which are effectively bosonic in character, 
lead to strong deviations  from  the above frequency dependence of 
${\rm Im} \chi^{-1}(\omega) $\cite{ACN,JS_1}, in disagreement with INS 
experiments. Thus, ${\rm Im} \chi^{-1}(\omega) $ does not arise
from spatially varying
stripes  even though they  certainly affect the spin degrees of freedom and 
can cause the decay of these excitations.

Finally,  NMR and other magnetic measurements    
 on various cuprates   which provide strong support
for spatial inhomogeneities\cite{CHO,Bor,PCH,Hunt,MHJ,Haase}  suggest
mostly 
spatially varying {\em spin} degrees of freedom but not necessarily  a 
strong
 modulation of the charge background. 
 
\begin{figure} [t]
\begin{center}
\leavevmode
\epsfxsize=7.5cm
\epsffile{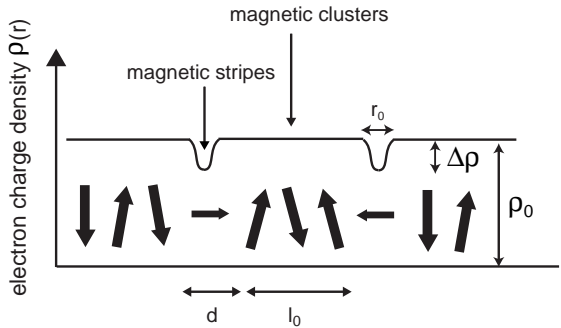}
\end{center}
\caption{ Schematic picture of charge and spin inhomogeneities.}
\label{mag_dom}
\end{figure}

The above points  suggest  that the inhomogeneities observed in 
the high-$T_c$ cuprates  might be  predominantly magnetic in 
origin with only moderate modulations of the charge density.
This is not unrealistic since in strongly
correlated systems small variations in the charge density can bring about 
substantial changes in the magnetic properties of the system. 
This scenario of a predominantly magnetic character of the inhomogeneities, 
leading to magnetic stripes and clusters, is
schematically presented in  Fig.~\ref{mag_dom}.  
Note, that the magnetic stripes, in 
which the electronic charge
density is lower than that in the magnetic clusters, represent
magnetic domain walls with a phase slip of $\pi$ in the ordering of the
spins.
Theoretically, we believe that this more homogeneous charge arrangement is a 
result
of quantum fluctuations beyond the mean field level investigated in 
Ref.~\cite{Voj99}.

The formation of magnetic clusters and stripes requires competing
interactions on different length scales. Besides the local  Coulomb
interaction, we need to identify a longer length scale interaction.  It
was  argued earlier that in the cuprate superconductors an increase of the 
magnetic correlation
length leads to strong vertex corrections, which in turn give rise to a
gradient coupling between the fermionic and spin degrees of freedom
\cite{vertex}. 
For a system with
an ordered ground state,  this takes the form of a dipole-dipole coupling
which has been
shown  to cause inhomogeneities and domain formation in
two dimensional magnetic systems.
The length scale of this gradient coupling should roughly
speaking be the magnetic correlation length, $\xi$. We thus need a
minimum value of $\xi$ to (a) generate the gradient coupling, and to (b)
observe the incommensuration. The temperature at which incommensuration
should appear is roughly set by $\xi(T) \approx l_0 = 4$, where the
last equality arises from the momentum position of the incommensurate
peaks. Since, as we argued earlier, $\xi(T_{cr}) \approx 2$, $T_{cr}$ thus
presents an upper bound for the formation of magnetic clusters. 
This anomalous coupling is of course enhanced once magnetic clusters are 
formed; the creation of magnetic inhomogeneities is a self-consistent process. 
The magnetic inhomogeneity scenario presented here is admittedly very 
qualitative 
and further microscopic investigations will be required to verify or disprove 
it.

\section{Conclusions}
\label{concl}
In this communication,
we investigated whether it is possible
to  understand theoretically INS and NMR data
within a single theoretical framework.
Based on the  observation of 
an incommensurate structure of the magnetic response by 
INS experiments~\cite{Dai98,AMH97},
 we considered whether this reflects a locally or globally
 incommensurate ordering. 
On analyzing NMR data on YBa$_2$Cu$_4$O$_{8}$ we found with the condition of 
$z=1$ 
scaling that a 
homogeneous incommensuration
is, within the framework given by Eq.(\ref{chi}),  inconsistent with 
the available
experimental data.
 We thus concluded that  the local magnetic structure as probed by NMR 
is likely commensurate.

We then investigated the effect of lattice corrections on both 
the parameter sets extracted from NMR data and 
on the local susceptibility measured
 in INS experiments.  We found that while the 
resulting corrections to the NMR parameters are only weak, the most 
pronounced effect of such corrections appears in the local 
susceptibility determined in INS measurements at frequencies above 
$\omega_{sf}$. 
Specifically, we found that decreasing the momentum cut-off leads 
to a suppression of $\chi''_{loc}$ 
at higher frequencies, thus 
improving the agreement with the 
experimental data. We argued that the 
remaining discrepancies 
can be explained by an experimental underestimate 
of the INS intensity at higher frequencies.  
Furthermore, we quantitatively compared INS and NMR data and found agreement
within a factor of 2.
  Given the large uncertainties in resolving $\chi''$ from the large
background and  in determining the absolute intensity of $\chi''$,
we believe that this result demonstrates  
that a consistent description of INS and NMR data can be obtained using 
the expression for $\chi$ given in Eq.(\ref{chi}).

Finally, we  discussed a spin and charge inhomogeneity scenario
to reconcile the local
commensurations, as seen in NMR experiments, and
the incommensurate peaks, seen by INS.
This scenario, even though similar in spirit to earlier
charge stripe pictures,  is based on a moderate to weak modulation
 of the charge density,
causing a pronounced inhomogeneity
in the magnetic properties and leading to magnetically coupled clusters
 separated by weakly correlated stripes.

We would like to thank A.V. Chubukov, P. Dai, B. Keimer, T. Mason, 
H. Mook,  R. Stern, C.P. Slichter and B. Stojkovic for valuable discussions.
 This work has been supported in part by the Science and Technology
 Center for Superconductivity through NSF-grant DMR91-20000 (D.K.M.), 
and by DOE at Los Alamos (D.P.).


\begin{thebibliography}{99}
\bibitem{FN1}  Due to the absence of a corresponding  elastic Bragg peak
 in neutron and X-ray scattering, stripes,
 if present at all in YBa$_2$Cu$_3$O$_{6+x}$, have to be dynamical in 
character.
\end{thebibliography}
\end{document}